# Photonic dark matter portal and quantum physics

S. A. Alavi, F. S. Kazemian

*Department of Physics, Hakim Sabzevari University, P. O. Box 397, Sabzevar, Iran.*

s.alavi@hsu.ac.ir; *alaviag@gmail.com*

**Abstract:** We study a model of dark matter in which the hidden sector interacts with standard model particles via a hidden photonic portal. We investigate the effects of this new interaction on the hydrogen atom, including the Stark, Zeeman and hyperfine effects. Using the accuracy of the measurement of energy, we obtain an upper bound for the coupling constant of the model as $f \leq 10^{-12}$. We also calculate the contribution from the hidden photonic portal to the anomalous magnetic moment of the muon as $a_\mu \leq 2.2 \times 10^{-23}$ (for the dark particle mass scale 100 MeV), which provides an important probe of physics beyond the standard model.

## 1 Introduction

There are various lines of evidence from cosmology and astrophysics for the existence of dark matter (DM). However, the nature of DM and its properties remain one of the most important problems in physics [1-11]. Attempting to answer the question of how we can find conclusive evidence for dark matter through direct detection experiments is of great importance to physicists.

The interaction of dark matter particles with standard model (SM) particles is not known. One possibility is that DM couples to the visible sector through the Higgs portal [12-14]. Another possibility is coupling through the photonic portal (sector) [15-19]. Coupling through the axion, neutrino and vector portals have also been proposed [20-22]. Since the particle model of dark matter is as yet unknown, it is worth studying all the possibilities and their consequences. In this paper we study the photonic DM portal and its consequences for quantum physics.

## 2 The model

In [15,19], the author proposed a model of a hidden sector of the universe consisting of sterile spin-1/2 fermions ("sterinos") and sterile spin-0 bosons ("sterons") interacting weakly through the mediation of sterile quanta of an antisymmetric-tensor field $A_{\mu\nu}$ ("A bosons"), but with a more intense coupling than through universal gravity. It is assumed that these sterile particles of the hidden sector can communicate with the SM sector through the electromagnetic field $F_{\mu\nu} = \partial_\mu A_\nu - \partial_\nu A_\mu$ weakly coupled to the A-boson field $A_{\mu\nu}$[16-18]:

$$L = \frac{1}{2}\sqrt{f}\,(\varphi F_{\mu\nu} + \xi\bar{\psi}\sigma_{\mu\nu}\psi)A^{\mu\nu}. \tag{1}$$

$\sqrt{f}$ and $\sqrt{f}\xi$ are small coupling constants. It is supposed that $\varphi = <\varphi>_{vac} + \varphi_{ph}$ with a spontaneously nonzero vacuum expectation value $\langle\varphi\rangle_{vac} \neq 0$. This coupling of photons to the hidden sector is called the "photonic portal".
The A-boson kinetic and SM electromagnetic Lagrangians together with Lagrangian (1) lead to the following field equations for $F_{\mu\nu}$ and $A_{\mu\nu}$ [15]:

$$\partial^\nu (F_{\mu\nu} + \sqrt{f}\varphi A_{\mu\nu}) = -j_\mu \tag{2}$$

$$(\partial_\mu \partial^\mu - \widetilde{M}^2)\partial^\nu A_{\mu\nu}^{(vac)} = \sqrt{f} <\varphi>_{vac} j_\mu, \tag{3}$$

where $\widetilde{M}^2 = M^2 + f<\varphi>_{vac}$, and $j_\mu$ and $M$ are the SM electromagnetic current and the mass scale of the A-boson respectively. The field equation (2) is the modified Maxwell's equations in the presence of the hidden sector interacting with the SM sector through the photonic portal. For a SM point charge at rest at $\vec{x'}$, $j_\mu(\vec{x}) = g_{\mu 0} e' \delta^{(3)}(\vec{x} - \vec{x'})$ and from Eq.(3) we have [19] :

$$\partial^\nu A_{\mu\nu}^{(vac)}(x) = -g_{\mu 0}\sqrt{f} <\varphi>_{vac} \frac{e'}{4\pi|\vec{x}-\vec{x'}|} e^{-\widetilde{M}|\vec{x}-\vec{x'}|}. \tag{4}$$

Using Eq.(4), it can be shown that the hidden sector correction $\delta V^{vac}$ to the Coulomb potential $V = \frac{ee'}{4\pi|\vec{x}-\vec{x'}|}$ is given by the following expression [19] :

$$\delta V^{(vac)} = \frac{f\langle\varphi\rangle_{vac}^2}{\widetilde{M}^2} \frac{ee'}{4\pi|\vec{x}-\vec{x'}|}\left(1 - e^{-\widetilde{M}|\vec{x}-\vec{x'}|}\right) + O(f^2). \tag{5}$$

## 3 Effects of a hidden photonic portal on the hydrogen atom

Hydrogen is the simplest atom to study, but much of what we know about the hydrogen atom can be extended to single-electron ions. It is also an ideal system for performing precise comparisons of theory and experiment. In this section, we therefore study the effects of the hidden photonic portal on the hydrogen atom.
Without giving any reason, the author in Ref. [19] considers a value for the coupling $f$ of 0.0917. In what follows we show that the coupling constant of the model should be much smaller than this value. It is also proposed in [19] that $\langle\varphi\rangle_{vac}^2 \sim (10^{-2} - 1)M^2$ and $M \sim (75 - 770)$ GeV, so $\frac{f\langle\varphi\rangle_{vac}^2}{\widetilde{M}^2} \sim 0.00092$, which is sufficiently small to allow us to use perturbation theory to find the energy shift due to the potential $\delta V^{(vac)}$ given by Eq.(5). To first order, the shift of energy is as follows:

$$\Delta E = <nlm|\delta V^{(vac)}|nlm> = -\frac{f\langle\varphi\rangle^2_{vac}}{\widetilde{M}^2}\frac{e^2}{4\pi}\left[<nlm\left|\frac{1}{r}\right|nlm> - <nlm\left|\frac{e^{-\widetilde{M}r}}{r}\right|nlm>\right], \qquad (6)$$

where $e$ is the electron charge, and $r = |\vec{x} - \vec{x'}|$, in which $\vec{x}$ and $\vec{x'}$ are the coordinate vectors of the electron and proton respectively.

The first term in Eq.(6) is the expectation value of $\frac{1}{r}$, which can be found in the quantum mechanics textbooks. The second term is the Yukawa-type potential and its expectation values are calculated in [23]. Alternatively, one can use Mathematica software [24] to calculate these expectation values and the integrals which appear in the following sections.

The energy corrections for the ground and one of the first excited states are as follows:

$$\Delta E_{100} = 3.43 \times 10^{-3} \, \text{eV}$$

$$\Delta E_{200} = 8.57 \times 10^{-4} \, \text{eV}$$

The accuracy of the energy measurement is $10^{-12}$ eV [25] and the values in Eq.(6) are large enough to be measured experimentally, but so far there is no evidence for these energy corrections in the hydrogen atom. One can therefore find an upper bound for the coupling $f$ using the fact that the energy corrections must be smaller than the accuracy of energy measurement, giving the result:

$$f \leq 10^{-12}$$

Apart from calculations and from the point of view of intuition, this value for $f$ is consistent with the fact that the coupling of the DM with luminescent (ordinary) matter is weak, but the value proposed in [15-19] seems to be too large to be consistent with this fact.

## 4 Contributions to the electron and muon magnetic moment from hidden photonic portal

The theoretical value for the muon anomalous magnetic moment $a_\mu$ is very important to study new phenomena in particle physics. At the tree level, the anomalous magnetic moment of the muon $g$ factor has a value of 2 and radiative contributions give some corrections to it. Hence it can be quantified as $a_\mu = \frac{(g_\mu - 2)}{2}$. Many years ago, a discrepancy was found between the experimentally measured value of the muon magnetic moment and the theoretically calculated value. The difference between $a_\mu^{SM}$ and $a_\mu^{Exp}$ (for the present status of the problem, see Section 7) has motivated studies of new physics scenarios that could offer an explanation, including supersymmetry, extra dimensions and hidden sectors, see [26] and references therein.

In this section we study the corrections due to the hidden photonic portal interaction on the magnetic moments of the electron and muon. To do this, we consider the related process which is described by Lagrangian (1) and represented by the Feynman graph in Figure 1.

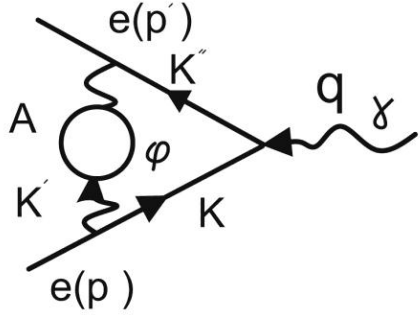

Figure 1. The dgiagram of related process to the correction described by Lagrangian in Eq.(1) on the magnetic moment of electron (muon).

This diagram corresponds to the following modified photon propagator:

$$\frac{-ig_{\mu\nu}}{k'^2} \to \frac{-i}{k'^2} \Pi_{\mu\nu} \frac{-i}{k'^2},$$

where :

$$\Pi_{\mu\nu} = (k'^2 g_{\mu\nu} - k'_\mu k'_\nu) \Pi^{A\varphi}(k'^2),$$

and :

$$\Pi^{A\varphi}(k'^2) = \frac{-f}{M^2} \int_0^1 z(1-z)dz \frac{-1}{(4\pi)^{d/2}} \frac{i}{2} \frac{\Gamma(n-d/2)}{\Gamma(n)} (\frac{1}{\Delta})^{n-d/2} + \frac{f}{M^2} \int_0^1 z^2 dz \frac{1}{(4\pi)^{d/2}} \frac{i\Gamma(n-d/2)}{\Gamma(n)} (\frac{1}{\Delta})^{n-d/2},$$

where $\Delta = -k'^2 z(1-z) + M^2$, so we have the following expression for the modified photon propagator :

$$\frac{-ig_{\mu\nu}}{k'^2} \to i(g_{\mu\nu} - \frac{k'_\mu k'_\nu}{k'^2}) \frac{\Pi^{A\varphi}(k'^2)}{k'^2}.$$

The photon propagator always occurs coupled to conserved currents and it follows from current conservation that the second term, which is proportional to the photon momentum, gives a vanishing contribution. So we remain with the calculation of the first term:

$$\frac{\Pi^{A\varphi}(k'^2)}{k'^2} = \int_0^\infty \frac{dt}{t} \frac{1}{t-k'^2} \frac{1}{\pi} \operatorname{Im}\Pi^{A\varphi}(t) ,$$

where $\frac{1}{\pi}\operatorname{Im}\Pi^{A\varphi}(t)$ is the spectral function [27] and is related to the decay rate through the following expression (see the illustration in Figure 2):

$$\Gamma = \sqrt{t}\operatorname{Im}\Pi^{A\varphi}(t) \tag{7}$$

This is the optical theorem and shows how the imaginary part of the spectral function is related to the decay rate.

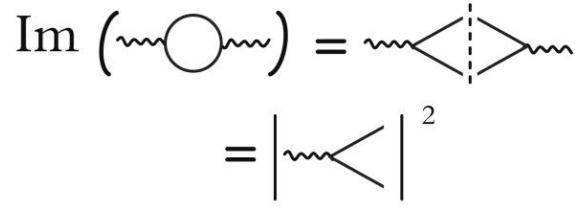

Figure 2. Diagrammatic relation between the spectral function and the decay rate in equation (7)

The decay rate for the process $A \to \varphi\gamma$ is given in [28]:

$$\Gamma(A \to \varphi\gamma) = \frac{f}{96\pi}\sqrt{t}(1-\frac{m_\varphi^2}{t})^3 , \tag{8}$$

where $m_\varphi$ is the mass of the steron and $t$ is the Mandelstam variable. From Eqs. (7) and (8) we have:

$$\frac{1}{\pi}\operatorname{Im}\Pi^{A\varphi}(t) = \frac{f}{96\pi^2}(1-\frac{m_\varphi^2}{t})^3 \theta(t-M^2) .$$

We also have :

$$a = \frac{1}{2}(g_\mu - 2) = \frac{1}{\pi}\int_0^\infty \frac{dt}{t} \mathrm{Im}\Pi^{A\phi}(t) k_e^2(t)$$

$$k_e^2(t) = (\frac{\alpha}{\pi})\int_0^1 dz \frac{z^2(1-z)}{z^2 + \frac{t}{m_e^2}(1-z)}$$

By integrating over z, for $k_\mu^2(t)$ we have $k_e^2(t) = \frac{1}{3}\frac{m_e^2}{t}$.

In the same way one can calculate the correction due to the hidden photonic portal interaction on magnetic moment of the muon. The results are given by the following expressions :

$$a_e = \frac{1}{3\times 96}\frac{f}{\pi}(\frac{\alpha}{\pi})\frac{m_e^2}{M^2} \qquad a_\mu = \frac{1}{3\times 96}\frac{f}{\pi}(\frac{\alpha}{\pi})\frac{m_\mu^2}{M^2}. \tag{9}$$

Using the upper bound for the coupling $f$ we found in Section 2, and taking $M = 100$ GeV, we obtain the following bounds for the electron and muon magnetic moments:

$$a_e \leq 0.5\times 10^{-27} \qquad a_\mu \leq 2.2\times 10^{-23}$$

The accuracy of the measurement of the muon magnetic moment is $10^{-11}$ [29], so the contributions to the muon magnetic moment due to a hidden photonic portal (with $M = 100$ GeV) is far from the present experimental sensitivity.

The interesting point is that if the dark matter mass is in the keV or MeV range, then its effects on quantum quantities such as the magnetic moments of the electron and muon and the energy levels of the hydrogen atom are detectable by present experimental techniques.

Recently there have been some signatures for light dark matter and some studies reveal that dark matter particles can have masses in the MeV and keV ranges [30-36], and ultra-light dark matter candidates can even have masses below the eV mass-scale [37], so it is not surprising that the dark particles in the model studied here can have masses in the keV range (see also Section 7). Therefore one can expect in future to be able to observe the effects of the hidden photonic portal on quantum systems such as those studied in this paper.

## 5 The permanent Stark and Zeeman effects

The Zeeman effect is very important in applications such as nuclear magnetic resonance spectroscopy, electron spin resonance spectroscopy, magnetic resonance imaging (MRI) and Mössbaure spectroscopy. The Stark effect has also become an increasingly important part of atomic, molecular and optical physics. In this section we therefore investigate the effects of a hidden photonic portal on the Stark and Zeeman effects.

Eq.(5) gives the corrections due to the hidden photonic sector on the electric potential of a particle with charge $e$ when the particle is at rest relative to the hidden sector particle (DM candidate ). If the particle, for instance the proton in an atom, is not at rest relative to the hidden sector particle and moves with relative velocity $v$, then the corrections to the electric and magnetic fields of the proton due to the presence of the photonic hidden sector are as follows (see Appendix):

$$\delta E' = \frac{f\langle\varphi\rangle^2_{vac}}{\widetilde{M}^2}\gamma\frac{e}{4\pi}[\frac{1}{r'^3}(1-e^{-\widetilde{M}r'}) - \frac{-\widetilde{M}}{r'^2}e^{-\widetilde{M}r'}\left((x-vt)\hat{i}+y\hat{j}+z\hat{k}\right)]$$

$$\delta B' = -\frac{f\langle\varphi\rangle^2_{vac}}{\widetilde{M}^2}\gamma v\frac{e}{4\pi}[\frac{1}{r'^3}(1-e^{-\widetilde{M}r'}) - \frac{-\widetilde{M}}{r'^2}e^{-\widetilde{M}r'}(-z\hat{j}+y\hat{k})] \qquad (10)$$

where : $r'=[\gamma^2(x-vt)^2+y^2+z^2]^{\frac{1}{2}}$ and $\gamma=\frac{1}{\sqrt{1-v^2/c^2}}$ ,and we have taken the velocity $v$ in the $x$ direction . It is seen that if $f=0$ the corrections $\delta E'$ and $\delta B'$ vanish. One can derive the interaction Hamiltonian of the hydrogen atom with the electric field $\delta E'$, which we call permanent Stark effect because it is due to the presence of dark matter rather than due to applying an external electric field:

$$H_{Stark} = -e\left(\delta\overrightarrow{E'}\right).\vec{r} = -\frac{f<\varphi>^2_{vac}}{\widetilde{M}^2}\frac{e^2}{4\pi}\gamma\left[\frac{(1-e^{-\widetilde{M}r'})}{r'^3} - \frac{\widetilde{M}e^{-\widetilde{M}r'}}{r'^2}\right][(x-vt)x+y^2+z^2]. \qquad (11)$$

We use time dependent perturbation theory (see Appendix) to find the transition probabilities. The results are shown in Table 1.

| $V=\frac{v}{c}$ | 0.1 | 0.2 | 0.3 | 0.4 | 0.5 |
|---|---|---|---|---|---|
| $P_1$ | 5.92707×10⁻⁶ | 3.09869×10⁻⁶ | 1.99511×10⁻⁶ | 1.76751×10⁻⁶ | 1.1498×10⁻⁶ |
| $P_2$ | 2.883×10⁻⁸ | 1.465×10⁻⁸ | 7.663×10⁻⁹ | 9.298×10⁻⁹ | 7.986×10⁻⁹ |
| V | 0.91 | | 0.92 | | 0.93 |
| $P_3$ | 5.49×10⁻⁷ | | 5.91×10⁻⁷ | | 4.55×10⁻⁷ |
| $P_4$ | 3.623×10⁻⁹ | | 3.788×10⁻⁹ | | 2.64×10⁻⁹ |

Table1. The nonrelativistic limit for the transition probabilities $P_1(\psi(100)\to\psi(200))$ , $P_2(\psi(100)\to\psi(211))$ and the relativistic cases $P_3(\psi(100)\to\psi(200))$, $P_4(\psi(100)\to\psi(211))$.

We observe that the transition probabilities are negligibly small (for $M$ in the GeV range), so we conclude that the hidden photonic portal (with sterile particles mass in the GeV range) can not cause the hydrogen atom to make a transition to the upper energy states.
The Hamiltonian of the Zeeman effect due to the presence of the photonic hidden sector is:
$$H = -\vec{\mu}_e.\overrightarrow{\delta B'}$$

Where $\overrightarrow{\mu_e} = \frac{e}{2mc}(\vec{L}+\overrightarrow{2S}) = \frac{e}{2mc}(\vec{J}+\vec{S})$, then we have :

$$H_{zeeman} = -\frac{f\langle\varphi\rangle_{vac}^2}{\widetilde{M}^2}\gamma v \frac{e^2}{4\pi}\frac{1}{2m_e c}[\frac{1}{r'^3}(1-e^{-\widetilde{M}r'}) - \frac{-\widetilde{M}}{r'^2}e^{-\widetilde{M}r'}][-z(J_y+S_y)+y(J_z+S_z)]. \tag{12}$$

This is also a time dependent Hamiltonian. Using the wave function of the unperturbed Hamiltonian $\psi_{m_j}^{j=l\pm\frac{1}{2}}$ one can calculate the transition probabities (see the Appendix). We found that the transition probabilities from the ground state to the $2s$ and $2p$ states for different velocities are zero.

## 6 Effect of a photonic portal on the hyperfine effect

In this section we study the magnetic influence of nuclear spin on electron spin. As was shown in Section 3, there is a correction due to the hidden photonic portal on the g factor of the electron. In what follows, we calculate the contribution of this correction to the hyperfine effect. The Hamiltonian is given by:

$$H = -\vec{\mu}_e \cdot \vec{B} \tag{13}$$

where $\vec{B} = \nabla \times \vec{A}$ is the magnetic field and the vector potential due to a point-like magnetic dipole is given by:

$$\vec{A}(\vec{r}) = \frac{-1}{4\pi}(\vec{M}\times\vec{\nabla})\frac{1}{r}. \tag{14}$$

Here $\vec{M} = \frac{Zeg_N}{2M_N c}\vec{I}$ is the magnetic dipole moment of the nucleus and $Ze$, $M_N$, $I$ and $g_N$ are its electric charge, mass, spin and gyromagnetic ratio respectively. After some calculations, Eqs. (13) and (14) give the following expression for the Hamiltonian:

$$H = -\frac{2}{3}\vec{\mu}\cdot\vec{M}\delta(\vec{r})$$

where $\vec{\mu}_e = \left(\frac{e}{m_e c}(1+\frac{\alpha}{2\pi}+\frac{1}{3\times 96\pi}\frac{f}{\pi}\left(\frac{\alpha}{\pi}\right)\frac{m_e^2}{M^2})\right)$. Using this Hamiltonian one can calculate the energy correction due to the photonic portal on the hyperfine effect. For the case of $l=0$ we have:

$$\Delta E = \frac{-8Zeg_n}{6M_n}\frac{1}{(na)^3}(\frac{e}{m_e c}(1+\frac{\alpha}{2\pi}+\frac{1}{3\times 96\pi}\frac{f}{\pi}(\frac{\alpha}{\pi})\frac{m_e^2}{M^2})\frac{1}{2}(F^2-S^2-I^2).$$

F is the total spin of the electron and the nucleus and $a$ is the Bohr radius. For the special case $F=1\to F=0$ we have:

$$\Delta E(F=1\to F=0) = \frac{-8Zeg_n}{3M_n}\frac{1}{(na)^3}(\frac{e}{m_e c}(1+\frac{\alpha}{2\pi}+\frac{1}{3\times 96\pi}\frac{f}{\pi}(\frac{\alpha}{\pi})\frac{m_e^2}{M^2})).$$

The second term, which gives the contribution of the hidden photonic portal, has the value of $0.4 \times 10^{-29}$ eV (for $M = 100$ GeV), which is too small to be measurable by present day experimental instruments.

## 7 Discussion

### 7-1. Present theoretical and experimental status of the anomalous magnetic moment of the muon

The present estimates for theoretical contributions (see Refs. [38,39] and references therein) to $a_\mu$ and experimental values [40], are presented in Table 2.

| QED | 11 658 4719 $\times$ $10^{-11}$ |
|---|---|
| Electroweak | 153 $\times$ $10^{-11}$ |
| HVP (hadronic vaccume polarization) | 6851 $\times$ $10^{-11}$ |
| HLBL (hadronic light by light scattering) | 116 $\times$ $10^{-11}$ |
| Standard Model | 116 591 802 $\pm$ 49 $\times$ $10^{-11}$ (0.42 ppm) |
| BNL measurement | 116 592 089 $\pm$ 63 $\times$ $10^{-11}$ (0.54 ppm) |
| Discrepancy | ~3.6σ |
| New experiment to reach | 140 ppb |

Table 2. Discrepancy between the SM calculations and the experimental values.

There is apparently a discrepancy between the SM calculations and the experimental values. From Eq.(9), we observe that if the mass scale of the A-boson is in the range 100 keV-MeV, the contributions of the hidden photonic portal can be measurable in future experiments in Fermilab and JPARC.

### 7.2. Observational evidence for keV X-rays and hidden photonic portal

It is well known that DM, which makes up 85% of the universe, is invisible. Physicists are attempting [41] to detect it by turning it into photons, so the photonic portal studied in this paper can help the research community in this direction. For instance, it is believed that recently observed the 3.5 keV X-ray line signal in the spectrum of 73 galaxy clusters, reported by the XXM-Newton X-ray observatory [42-44], has a DM origin. The photonic hidden sector could be an appropriate candidate to provide an explanation for the observation (work in preparation), because the dark particles (with mass in the keV mass range) in this model can decay directly into X-ray photons with keV energy.

### 7.3. Extension of the model to a generic hidden photonic portal

Beside the present results it would be interesting to investigate if the study can be extended to a generic photonic portal DM so that our study can be applied to a more generic class of models. Extensions of the SM with an additional hidden U(1) gauge symmetry (hidden sector) have recently attracted much attention. In common extensions of the SM an extra U(1) is kinetically mixed with the electromagnetic/hypercharge U(1) of the SM (see e.g.,[45] and references therein), which corresponds to an extra photon-like particle, the hidden photon (HP). All SM particles are uncharged under this new gauge group, and hidden photons interact very weakly with ordinary photons via a so-called kinetic mixing :

$$L = \frac{k}{2} B_{\mu\nu} F^{\mu\nu}$$

where $F_{\mu\nu}$ and $B_{\mu\nu}$ are the photon ($A_\mu$) and HP ($B_\mu$) field strengths. The current bounds on the kinetic mixing parameter $k$ range from $10^{-13}$ to $10^{-3}$.

If DM is composed of hidden-sector photons that kinetically mix with photons of the visible sector, then one expects photon-HP oscillations (photon↔HP) which are similar to neutrino flavour oscillations. Phenomenologically, this model leads to a number of surprising phenomena : distortions of the CMB induced by the photon-HP mixing [46], astrophysical searches for a HP signal in the radio regime and radio sources [47], light-shining through walls [48] and production of HP in the early universe contributing to the dark radiation and DM of the universe [49,50].

One can therefore combine this model and the model discussed in our paper to construct a more generic model and try to explain unsolved phenomena.

## 8 Conclusions

Dark matter interaction with SM particles is as yet unknown. Depending on the particle physics model of DM, it could happen through : the axion, Higgs, neutrino, photonic or vector portal. In this paper we have studied the effects of the hidden photonic portal (sector) proposed in [15-19] on the hydrogen atom, including the Stark, Zeeman and hyperfine effects. We obtained an upper bound for the coupling constant of the model as $f \leq 10^{-12}$. We have also calculated the effects of the photonic hidden sector on the electron and muon magnetic moments. We find that if the mass scale of the dark matter particles is in the GeV range or higher, then corrections are not detectable with present day instruments. However, if it is in the keV or MeV range, then one can expect the corrections due to the photonic hidden sector to be detectable by present and near future experimental techniques.

**Appendix A:**

**Derivation of the electric and magnetic fields of the Stark and Zeeman effects.**

In the reference frame in which the hydrogen atom (proton) is at rest relative to the hidden sector particle, we have the following corrections on the scalar and vector potentials:

$$\delta\varphi = \frac{f\langle\varphi\rangle_{vac}^2}{\widetilde{M}^2} \frac{e}{4\pi r}\left(1 - e^{-\widetilde{M}r}\right)$$
$$\overrightarrow{\delta A} = 0 \qquad\qquad\qquad\qquad\qquad\qquad\qquad\qquad\qquad\qquad\qquad\qquad\text{(A1)}$$

The scalar and vector potentials are the components of the electromagnetic four-potential, so they transform according to the Lorentz transformations. Therefore in a reference frame in which the hydrogen atom moves with a constant velocity $\vec{v}$ with respect to the dark matter candidate, we will have the following corrections to the scalar and vector potentials:

$\delta\varphi' = \gamma\, \delta\varphi$

$\overrightarrow{\delta A'} = \gamma \vec{v} \delta\varphi$

Now, one can derive the electric and magnetic fields using the widely known relations :

$\overrightarrow{\delta E'} = -\nabla \overrightarrow{\delta\varphi'} - \frac{\partial(\overrightarrow{\delta A'})}{\partial t}$ , $\overrightarrow{\delta B'} = \nabla \times \overrightarrow{\delta A'}$ ; the results are presented in Eq.(10). As an another option one can first calculate $\delta E$ and $\delta B$ from Eq.(A1) and then find $\delta E'$ and $\delta B'$ using the well-known transformations of the parallel and perpendicular (to the relative velocity $\vec{v}$) components of the electric and magnetic fields :

$\delta E'_{\parallel} = \delta E_{\parallel}$ ;  $\delta E'_{\perp} = \gamma[\delta E_{\perp} + (\vec{v} \times \overrightarrow{\delta B})_{\perp}]$

$\delta B'_{\parallel} = \delta B_{\parallel}$ ;  $\delta B'_{\perp} = \gamma[\delta B_{\perp} - \frac{1}{c^2}(\vec{v} \times \overrightarrow{\delta E})_{\perp}]$ ,

which leads to the same results.

**Appendix B : Stark effect**

In time-dependent perturbation theory, the amplitude for the transition $|m'> \to |m>$ is given by :

$c^{(1)}_{mm'}(t) = \frac{-i}{\hbar}\int_0^\infty \langle m|H_{Stark}|m'\rangle = \frac{-i}{\hbar}\int_0^\infty e^{i\omega_{mm'}t}(H_{Stark})_{mm'}(t')dt'$

where $\omega_{mm'} = \frac{(E_m - E_{m'})}{\hbar} = -\frac{1}{2}\mu c^2 \frac{(Z\alpha)^2}{\hbar}\left(\frac{1}{m^2} - \frac{1}{m'^2}\right)$, μ is the reduced mass of the electron and proton and the probability of the transition is $P_{mm'}(t) = |c^{(1)}_{mm'}(t)|^2$. The time dependence of the perturbing Hamiltonian reveals itself through $(x - vt)x$ and $r' = [\gamma^2(x-vt)^2 + y^2 + z^2]^{\frac{1}{2}}$ in Eq.(11). Expressing $x, y$ and $z$ in spherical polar coordinates and using the wave functions of the ground state $|m'> \equiv |100>$ and exited states $|m> \equiv |nlm>$ , which can be found in the textbooks on quantum mechanics, one can calculate the integrals and the transition probability for the Stark effect produced by the hidden photonic portal in Section 4.

## Appendix C: Zeeman effect

The unperturbed Hamiltonian of the Zeeman effect is as follows :

$$H_0 = \frac{\vec{p}^2}{2\mu} - \frac{Ze^2}{r} + \frac{1}{2m_e^2 c^2} \frac{Ze^2}{r^3} \vec{L}.\vec{S}$$

For the perturbing Hamiltonian $H_{Zeeman}$ in Eq.(12) we have :

$$\left\langle \psi_{m_j}^{j=l+\frac{1}{2}} \middle| S_z \middle| \psi_{m_j}^{j=l+\frac{1}{2}} \right\rangle = \frac{\hbar m_j}{2l+1}; \qquad \left\langle \psi_{m_j}^{j=l-\frac{1}{2}} \middle| S_z \middle| \psi_{m_j}^{j=l-\frac{1}{2}} \right\rangle = -\frac{\hbar m_j}{2l+1}$$

$$\left\langle \psi_{m_j}^{j=l+\frac{1}{2}} \middle| S_y \middle| \psi_{m_j}^{j=l+\frac{1}{2}} \right\rangle = 0; \qquad \left\langle \psi_{m_j}^{j=l-\frac{1}{2}} \middle| S_y \middle| \psi_{m_j}^{j=l-\frac{1}{2}} \right\rangle = 0$$

$$\left\langle \psi_{m_j}^{j=l+\frac{1}{2}} \middle| J_y \middle| \psi_{m_j}^{j=l+\frac{1}{2}} \right\rangle = 0; \qquad \left\langle \psi_{m_j}^{j=l-\frac{1}{2}} \middle| J_y \middle| \psi_{m_j}^{j=l-\frac{1}{2}} \right\rangle = 0; \qquad \left\langle \psi_{m_j}^{j=l\pm\frac{1}{2}} \middle| J_z \middle| \psi_{m_j}^{j=l\pm\frac{1}{2}} \right\rangle = m_j \hbar$$

The rest of the calculations proceed as explained in Appendix B.

## Acknowledgement

S. A. Alavi would like to thank the Department of Physics of the University of Turin (Italy) for hospitality where some parts of this work were done. He is also very grateful to Rouzbeh Allahverdi (university of New Mexico, USA) for his valuable comments and Pere Masjuan (Johannes Gutenberg-Universität, Mainz, Germany) and Gernot Eichmann (Justus-Liebig-Universität, Giessen, Germany) for useful discussions.

## References

[1]. Z.Shou-Hua, Chin. Phys. Lett. 24(2007) 381.

[2]. S. C. Yoon, F. Iocco, S. Akiyama, Astrophys.J. 688 (2008) L1.

[3]. Y. Hosotani, P. Ko, M. Tanaka, Phys.Lett.B. 680 (2009) 179.

[4]. Andrzej M. Szelc, Acta Phys.Polon.B. 41 (2010) 1417.

[5]. J. Goodman, et.al., Nucl.Phys.B. 844 (2011) 55.

[6]. Q. Hong-Yi, W. Wen-Yu, X. Zhao-Hua, Chin. Phys. Lett. 28 (2011) 111202.

[7]. Q. Hong-Yi, W. Wen-Yu, X. Zhao-Hua, CMS Collaboration, JHEP 09 (2012) 094.

[8]. L. Xu, Y. Chang, Phys. Rev. D. 88 (2013) 127301.

[9]. A. Crivellin, F. D'Eramo, M. Procura, Phys. Rev. Lett. 112 (2014) 191304.

[10]. L. Carpenter, et.al., Phys. Rev. D. 89 (2014) 075017.

[11]. K. Freese, arXiv:1501.02394, invited review for Reports on Progress in Physics.

[12]. J. March-Russell, S. M. West, D. Cumberbatch and D. Hooper, JHEP 7 (2008) 058.

[13]. C. Englert, T. Plehn, D. Zerwas and P. M. Zerwas, Phys. Lett. B. 703 (2011) 298.

[14]. L. Lopez-Honorez, T. Schwetza, J. Zupan, Phys. Lett. B. 716 (2012) 179.

[15]. W. Królikowski, arXiv: 1008.4756.

[16]. W. Królikowski, Acta Phys. Polon. B 39 (2008)1881.

[17]. W. Królikowski, Acta. Phys. Polon. B. 40 (2009) 111.

[18]. W. Królikowski, Acta . Phys. Polon. B. 40 (2009) 2767.

[19]. W. Królikowski, Acta. Phys. Polon. B. 42 (2011) 1261.

[20]. Y. Nomura, J. Thaler, Phys. Rev. D. 79 (2009) 075008.

[21]. J. Cherry, A. Friedland, I. M. Shoemaker, arXiv:1411.1071.

[22]. Jiang-Hao Yu, Phys. Rev. D 90, 095010 (2014).

[23]. M. Hamzavi , et.al., Chin. Phys. Lett. 21 (2012) 80302.

[24]. Wolfram Mathematica, The world's definitive system for modern technical computing ; http://www.wolfram.com/mathematica/.

[25]. L. Essen et.al., Nature 229 (1971).

[26] D. McKeen, Annals Phys. 326 (2011) 1501.

[27]. J. P. Miller, E. de Rafael and B. Lee Roberts, Rept. Prog. Phys. 70 (2007) 795.

[28]. W. Królikowski, arXiv: 0903.5163.

[29]. K. Olive, *et al.*, Review of Particle Physics, Chin. Phys. C 38 (2014) 090001.

[30]. P. deNiverville, D. McKeen, A. Ritz, Phys. Rev. D. 86 (2012) 035022.

[31], P. deNiverville, M. Pospelov, A. Ritz, Phys. Rev. D 84, 075020 (2011).

[32]. B. Batell, R. Essig, Z. Surujon, Phys. Rev. Lett. 113 (2014) 171802.

[33]. B. Batell, M. Pospelov, and A. Ritz, Phys. Rev. D 80 (2009) 095024 (2009).

[34]. R. Essig et.al., JHEP11(2013)167.

[35]. K. N. Abazajian, Physics 7 (2014) 128.

[36]. A. Merle, A. Schneider, Report number: MPP-2014-348**,** arXiv:1409.6311.

[37]. Babette Döbrich, arXiv:1501.03274, Report number : DESY 15-003, contribution to the 24th European

Cosmic Ray Symposium.

[38]. P. Masjuan, Nuclear Physics B Proceedings Supplement 00 (2014) 1.

[39]. G. Eichmann et.al., arXiv :1411.7876.

[40]. S. Henry(university of Oxford), "News From Fermilab", https://indico.in2p3.fr/event/10304/contribution/20/material/slides/. December 9, 2014.

[41]. M. R. Francis, Symmetry, A joint Fermilab/SLAC publication, January 2015.

[42]. E. Bulbul, et.al., Astrophys. J. 789 (2014) 13.

[43]. A. Boyarsky, et.al., Phys. Rev. Lett. 113, 251301 (2014).

[44]. XMM-Newton Science Analysis System, http://xmm.esa.int/sas/.

[45]. Javier Redondo, MPP-2014-33, arXiv:1501.07292.

[46]. A. Mirizzi, J. Redondo and G. Sigl, JCAP 0903 (2009) 026.

[47]. A. P. Lobanov, H. -S. Zechlin and D. Horns, Phys. Rev. D 87 (2013) 6 065004.

[48]. J. Redondo and A. Ringwald, Contemp. Phys. 52 (2011) 211.

[49]. H. Vogel and J. Redondo, JCAP 1402 (2014) 029.

[50]. J. Redondo and M. Postma, JCAP 0902 (2009) 005.